\documentclass[preprint,amsmath,floats,showpacs,aps,superscriptaddress,12pt]{revtex4}

\usepackage[english]{babel}

\usepackage{graphicx}

\begin{document}

\title{Principles of wide bandwidth acoustic detectors and the  single-mass DUAL detector}
\author{Michele Bonaldi}
\email[Corresponding author: ]{bonaldi@science.unitn.it}
\affiliation{Istituto di Fotonica e Nanotecnologie CNR-ITC and INFN Trento, I-38050 Povo (Trento), Italy}
\author{Massimo Cerdonio } \affiliation{INFN Padova Section and Department of Physics University of Padova, via Marzolo 8, I-35100 Padova, Italy}
\author{Livia Conti} \affiliation{INFN Padova Section and Department of Physics University of Padova, via Marzolo 8, I-35100 Padova, Italy}
\author{Paolo Falferi}\affiliation{Istituto di Fotonica e Nanotecnologie CNR-ITC and INFN Trento, I-38050 Povo (Trento), Italy}
\author{Paola Leaci} \affiliation{Department of Physics University of Trento and INFN Trento, I-38050 Povo (Trento), Italy}
\author{Stefano Odorizzi} \affiliation{Enginsoft s.p.a., Via Giambellino, 7, 35129 Padova, Italy}
\author{Giovanni A. Prodi} \affiliation{Department of Physics University of Trento and INFN Trento, I-38050 Povo (Trento), Italy}
\author{Mario Saraceni} \affiliation{Enginsoft s.p.a., Via Giambellino, 7, 35129 Padova, Italy}
\author{Enrico Serra} \affiliation{Istituto Trentino di Cultura ITC-IRST Microsystem Division, 38050 Povo (Trento), Italy}
\author{Jean Pierre Zendri } \affiliation{INFN Padova Section, via Marzolo 8, I-35100 Padova,
Italy}
\date{\today}

\begin{abstract}
We apply the standard theory of the elastic body to obtain a set of equations describing the behavior of an acoustic Gravitational Wave detector, fully taking into account the 3-dimensional properties of the mass, the readout and the signal. We show that the advantages given by a Dual detector made by two nested oscillators can also be obtained by monitoring two different acoustic modes of the same oscillator, thus easing the detector realization. We apply these concepts and by means of an optimization process we derive the main figures of such a single-mass Dual detector designed specifically for the frequency interval $2-5\;$kHz. Finally we calculate the SQL sensitivity of this detector. 
\end{abstract}

\maketitle

\section{Introduction}
\label{sec:intro}

Direct detection of gravitational waves (GW), for physics and astrophysics study, is one of the great challenges of contemporary experimental physics. Resonant mass  bar  detectors \cite{louisiana1,auriga1,rog1}, the first historically to come to continuous operation, have been improved by 4 orders of magnitude in energy sensitivity during the last 40 years, so that they can detect energy changes of a 2300 kg bar as little as a few hundred of quanta of vibration at about 1 kHz with a bandwidth up to 100 Hz \cite{rog2,auriga2}. 

Even though the direct and unambiguous detection of GW remains an important goal in today's experimental physics, the scientific interest is moving toward the possibility of studying the physical and astrophysical features of the radiation. In fact gravitational waves will yield unique information about the coherent bulk motions of  matter generating the radiation, revealing features of their source which could not be investigated by electromagnetic, cosmic ray, or neutrino studies  \cite{Thorne_300y}.
In order for acoustic detectors to take part to this \textquotedblleft observatory phase\textquotedblright, a substantial improvement should be achieved both in terms of absolute sensitivity and detection bandwidth. In fact only a detector sensitive in a wide frequency range, with a good sky coverage for most of the time, will permit to follow the star system evolution during a long time interval, improving the signal reconstruction and the detection probabilities. 
Wideband acoustic detectors were recently proposed  and their expected sensitivity analytically evaluated in two ideal configurations \cite{cerdonio1,bonaldi1};
a precise analytical evaluation of the limit sensitivity  of such detectors is still lacking, due to the complexity of the problem. 

Present acoustic gravitational wave detectors consist of a test mass, which is strained under the effect of the gravitational wave. A position meter monitors a relevant dimension of the test mass, and thereby deduces the equivalent force signal. The sensitivity evaluation are traditionally made by modeling the test mass as a simple mechanical oscillator with a properly chosen effective mass. This approximation is very good when a narrow band detector sensitive around one of its internal resonances is considered: the detector behavior within its bandwidth is essentially determined by this resonant mode. On the contrary this approximation is seriously limited when applied to a wideband detector. In this case the detector is also sensitive at frequencies where many internal resonant modes of the test mass will equally contribute to the output signal, each with different amplitudes and phases according to their resonant frequencies. The resulting test mass dimension variation is then obtained as the sum of a number of contributions, and can be enhanced or reduced by the phase relation of the internal resonant modes. 

Acoustic detectors work at or between the resonant frequencies of the  lowest quadrupolar normal modes of the test mass, which are set by the geometry and material properties. GW sensitive modes in a practical test mass (up to a few meters in diameter) range from about 1 to several kHz. As a general rule large test masses give better sensitivities and lower frequencies, at the expenses of a greater realization difficulty. The current predictions of the expected rates of GW signals in the kHz band  are in rapid evolution, and a full discussion of this topic goes beyond the sake of this paper. We point out that a variety of sources have been considered recently to give possibly significant GW emission in the acoustic frequency range. As an example we give the design guidelines for a detector sensitive in the 2-5 kHz range: it would observe a number of interesting phenomena \cite{shibata,pretorius,campanelli} in a frequency range not easily covered by interferometric detectors due to the laser shot noise contribution \cite{saulsonlibro}. 

In Section \ref{section:SQLmodel} we develop the general mathematical framework of the complete test mass + readout system, leading to a set of formulas in which the response of the full detector is related to an external GW  force field and to the readout back action force. We also evaluate the Standard Quantum Limit (SQL) sensitivity of the detector in the ideal case and propose an optimization strategy accounting for  the presence of antiresonance frequencies in the transfer functions and the noise properties of the readout. Then in section \ref{section:singledual} we calculate the  SQL sensitivity of a single-mass Dual acoustic detector specifically designed to be sensitive in a broad frequency range. By an in-depth study of the phase relations between the acoustic modes, we obtain the geometrical conditions which optimize the detector sensitivity. As our calculation takes into account only plane strain resonant modes of the test mass, we validate the results by a Finite Element Method (FEM) simulation of the full 3-dimensional body.
Conclusions and limits of the current work are summarized in Section \ref{sec:conclusions}.

\section{Wideband resonant GW detector general model}
\label{section:SQLmodel}

The detector consists of a test mass and a readout system, which monitors a properly chosen physical observable of the test mass, namely an average of the surface displacement over a sensitive area. Provided that all external noise sources are reduced down to a negligible level, the test mass changes its dimensions due to the combined effects of the external GW force, the back action force  from the readout and the thermal noise force. In contrast with the lumped model case (which holds for a narrow band resonant detector), the effects of these forces need to be evaluated differently when the detector is considered as an elastic body. In fact the GW equivalent force is applied on the volume of the body, while the back action force is applied on the body surface. A complete description of the detector  requires the knowledge of  the position and extension of the surfaces on the test mass which the readout samples. In the most general case the readout sensitivity will be position dependent, and a weight function must be defined in order to fully describe the interaction with the test mass.
 
In the following, vectors are indicated by  bold characters.

\subsection{Test mass mode expansion}
\label{subsection:testmassexpansion}
 The equations  of the motion  of an elastic body of density $\rho$,
forced by a force density $\textbf{F}(\textbf{r},t)$, can be summarized as \cite{Love}:
\begin{equation}
\label{eq:start}
 \rho\frac{\partial^2 \textbf{u}(\textbf{r},t)}{\partial
t^2}-\emph{L}\,[\textbf{u}(\textbf{r},t)]= \textbf{F} (\textbf{r},t)\,,
\end{equation}
with the appropriate initial and boundary conditions. Here $\textbf{u}(\textbf{r},t)$ is the displacement field of the elastic body and $\emph{L}\,[\textbf{u}(\textbf{r},t)]$ is defined as:
\begin{equation}
\emph{L}\,[\textbf{u}]\,=(\lambda + \mu)\:\nabla (\nabla \cdot
\textbf{u})  + \mu\,\nabla^2 \textbf{u}\,.
\end{equation}
The Lam\`{e} coefficients $\lambda$ and $\mu$ depend on the
Poisson ratio $\sigma_p$ and on the Young modulus $Y$ of the
material:
\begin{equation}
\label{eq:lame}
 \lambda=
\frac{Y\sigma_p}{(1+\sigma_p)(1-2\sigma_p)} \qquad
\mu=\frac{Y}{2(1+\sigma_p)}\,.
\end{equation}
We define the displacement normal modes $\textbf{w} _{ n}(\textbf{r})$ as the
solutions of the eigenvalue problem:
\begin{equation}
-\rho\,\omega^2_{n}\,
\textbf{w}_{n}\,=\,\emph{L}[\textbf{w}_{n}] \,.
\label{eigenproblem0}
\end{equation}
with the boundary conditions on $\textbf{w}_{n}$ given by the requirement that components of the stress normal to the test mass surfaces vanish on the test mass surfaces. The normal modes constitute an orthogonal complete system, and can be normalized to satisfy the condition:
\begin{equation}
\int_V
\rho \, \textbf{w}_{n}(\textbf{r})\cdot\textbf{w}_m(\textbf{r})\,
\textit{d}V =M \,\delta_{n\,m}\:\
\label{ortho0}
\end{equation}
where the volume integral is performed on the test mass volume $V$ and $M$ is its total mass. 

The solution of Eq. (\ref{eq:start}) may then  be written as a superposition of the normal modes:
\begin{equation}
 \textbf{u}(\textbf{r},t)= \sum_{ n} \textbf{w}_{  n}(\textbf{r})\,q_{ n}(t)\, .
\label{sumsolution0}
\end{equation}
The sum runs over an infinite number of elements, as normal modes constitute an infinite enumerable set within the continuum mechanics mathematical framework \cite{Nmodes}. The functions $q_{n}(t)$ in  Eq. (\ref{sumsolution0}) represent the time development of the ($n$)-th mode, with initial value given by:
\begin{equation}
q_{ n}(0)=\frac{1}{M}\,\int_V\,\textit{d}V \;\rho\,\textbf{u}\, (\textbf{r},\,0)\cdot\textbf{w}_n(\textbf{r})\,.
\end{equation}
 The evaluation of the 
explicit form  of $q_{n}(t)$  is straightforward  \cite{meirovitch} when the driving force can be factorized as:
\begin{equation}
\label{eq:forcefactorize}
{F}(\textbf{r},\,t)=G_t(t)\,\textbf{G}_r(\textbf{r})\,.
\end{equation}
In fact  Eq. (\ref{eq:start}) becomes:
\begin{displaymath}
\rho\sum_{ n}
\textbf{w}_n(\textbf{r})\,\frac{\partial^2
q_n(t)}{\partial t^2}-\sum_{n}
q_n(t)\,\emph{L}[\textbf{w}_n(\textbf{r})]\,
=\,G_t(t)\,\textbf{G}_r(\textbf{r})\, ,
\end{displaymath}
and we can apply Eq. (\ref{eigenproblem0}),
 multiply by $\textbf{w}_m$ and integrate using Eq.
(\ref{ortho0}), obtaining:
\begin{displaymath}
M \,\frac{\partial^2 q_m(t)}{\partial t^2}+M\,\omega_m^2\,
q_m(t)\,
=\,G_t(t)\,\int_V\,\textit{d}V \textbf{G}_r(\textbf{r})\cdot\textbf{w}_m(\textbf{r})\,.
\end{displaymath}
The time development of the $(m)$-th mode is the same of a forced harmonic oscillator.
In the frequency domain (here and in the following we indicate the Fourier transform with a tilda) we have then:
\begin{equation}
\label{eq:qsolutionnodiss}
\widetilde{q}_m(\omega)\,=\,\frac{1}{M}\frac{\widetilde{G}_t(\omega)}{(\omega^2_{m}-\omega^2)}\,\int_V \textit{d}V\,\textbf{G}_r(\textbf{r})\cdot\textbf{w}_m(\textbf{r})\,
 .
\end{equation}
In the hypothesis of dissipation due to the material structure, the system losses can be modeled in the frequency domain by including explicitly a damping term \cite{saulson}. We have then: 
\begin{equation}
\label{eq:qsolution}
\widetilde{q}_m(\omega)\,=\,\frac{1}{M}\frac{\widetilde{G}_t(\omega)}{(\omega^2_{m}-\omega^2)+
i \omega^2_{m}
\phi_{m}(\omega)}\,\int_V \textit{d}V\,\textbf{G}_r(\textbf{r})\cdot\textbf{w}_m(\textbf{r})\,
 ,
\end{equation} 
The function $\phi_{n}(\omega)$ represents the phase lag between the mass displacement at a
given frequency and a monochromatic driving force. Experimentally, $\phi_{n}(\omega)$ is found to be roughly constant  (this is usually referred to as ``structural damping'') and, for low loss materials, it is $\phi_{n}(\omega) \ll 1$. It can be shown that the normal mode expansion is possible only if the damping term is homogeneous over the test mass volume, as any inhomogeneity of the structure damping causes a coupling between different modes \cite{yamamoto1}. When the losses are frequency independent the material quality factor $Q=1/\phi$ can be used equivalently in place of the phase lag.

\subsection{Readout}

To detect the external force we measure the resulting strain on some test mass surface $S$. In the small displacement approximation the observable physical quantity, $X$, of the system may be defined as:
\begin{equation}
X(t)= \int_S \textit{d} s\, \textbf{P}(\textbf{r})\,\cdot\,\textbf{u}(\textbf{r},t) 
  \,.
\label{observabledef}
\end{equation}
Here ${\textbf{P}(\textbf{r})} $ is a weighting function  and the integral is performed on the chosen surface $S$ of the test mass. We can decide freely the portions of the test mass surface which are sampled and their relative weight in the construction of our output variable $X(t)$: the spatial form of the weight function $\textbf{P}(\textbf{r})$ reflects our measurement strategy and detection scheme. For example, in an optical readout, $\textbf{P}(\textbf{r})$ is proportional to the beam spot power profile \cite{ottico}, while in a capacitive readout  \cite{capacitivo} we have $P(r)=1/S_0$ on the surface $S_0$ of the electrodes and  $P(r)=0$ outside (as border effects are usually negligible). 

By Eqs. (\ref{sumsolution0}, \ref{observabledef}) the observable
physical quantity of the system is:
\begin{equation}
X(t)= \,\sum_{n} q_n(t)\int_S \textit{d} s\,
\textbf{P}(\textbf{r}) \,\cdot
\,\textbf{w}_n(\textbf{r})\:,
\end{equation}
or, in the frequency domain:
\begin{eqnarray}
&&\widetilde{X}(\omega)=\label{genericresponse}\frac{\widetilde{G}_t(\omega)}{M}\\ &&
\sum_{n}\,\frac{\big[\int_V \textit{d}V\,\textbf{G}_r(\textbf{r})\cdot\textbf{w}_n(\textbf{r})\big]
\; \big[\int_S \textit{d}
s\,\textbf{P}(\textbf{r})\,\cdot
\, \textbf{w}_n(\textbf{r})\big]  \,}{(\omega^2_{n}-\omega^2)+ i \omega^2_{n}
\phi_{n}(\omega)}\,. \nonumber 
\end{eqnarray}

This equation allows to evaluate  the motion of the system when the specific form of the driving force spatial dependence and of the readout weight function are given. Then we evaluate the detector response when driven by the relevant forces: a GW,  the readout back action force and the thermal noise.

In the case of a linearly polarized gravitational wave propagating along the $z$-axis, the force density applied on the system may be written in cylindrical coordinates $(r,\theta)$ as \cite{misner}:
\begin{eqnarray}
\label{gwaveplus}
G_t(t)&=&\frac{1}{2} \,\rho \,\ddot{h}(t) \\
\textbf{G}_r(\textbf{r})&\equiv &\,\textbf{W}(\textbf{r})= {r }  \,\cos\,(\,2\,\theta+\psi\,)
\,\textbf{i}_r -  {r} \,\sin\,(\,2\,\theta+\psi\,)\,
\textbf{i}_\theta\,, \nonumber
\end{eqnarray}
where $h(t)$ is the metric perturbation associated to the GW, $\psi=0$ for the GW + polarization and $\psi=\pi/4$ for the $\times$ polarization. When we  substitute these in  Eq. (\ref{genericresponse}), the system response to a metric perturbation $h$ becomes:
\begin{equation}
\label{eq:gwHdef}
\widetilde{X} (\omega)=\,\widetilde{h}(\omega)\,H _{GW}(\omega)\:,
\end{equation}
where the detector output is fully described by the transfer function 
$H_{GW}(\omega)$:
\begin{eqnarray}
\label{gwtransfer}
&&H_{GW}  (\omega)= \frac{1}{2\,V} \,\\&& \sum_{n}
\,\frac{-\omega^2\:\big[\int_V \textit{d}V \, \textbf{W}(\textbf{r})\cdot\textbf{w}_n(\textbf{r})
\big]\; \big[\int_S\textit{d}s\, {\textbf{P}(\textbf{r})\,\cdot
\,\textbf{w}_n(\textbf{r})}  \big]}{(\omega^2_{n}-\omega^2)+ i \omega^2_{n}
\phi_{n}(\omega)} \,, \nonumber 
\end{eqnarray}

The second relevant force applied to the system is the readout back action. Being exerted by the readout, this force is  applied proportionally to the weight function $\textbf{P}(\textbf{r})$: 
\begin{equation}
\textbf{F}_{BA}(t,r)=F_{BA}(t) \,\textbf{P}(\textbf{r})\,.	
\label{eq:backaction}
\end{equation}

The test mass response to this input force can again be evaluated by Eq. (\ref{genericresponse}), if the back action surface force Eq. (\ref{eq:backaction}) is considered in place of the volume force field $G_t(t) \textbf{G}_r(r)$:
\begin{eqnarray}
\widetilde{X}(\omega)&=&T_{BA}(\omega)\,\widetilde{F}_{BA}(\omega)\nonumber\\
&=&\,\frac{\widetilde{F}_{BA}(\omega)}{M}\,\sum_{n}
\,\frac{ \big[\int_s \textit{d}s\,\textbf{P}(\textbf{r}) \,\cdot
\,\textbf{w}_n(\textbf{r}) \big]^2}{(\omega^2_{n}-\omega^2)+ i \omega^2_{n}
\phi_{n}(\omega)}\,\:. \label{backactionresponse}
\end{eqnarray}
The transfer function of the system relative to the back action force is then:
\begin{equation}
T_{BA}(\omega)=\,\frac{1}{M}\,\sum_{n}
\,\frac{ \big[\int_s\textit{d}s \,{\textbf{P}(\textbf{r})}\,\cdot
\,\textbf{w}_n(\textbf{r})    \big]^2}{(\omega^2_{n}-\omega^2)+ i \omega^2_{n}
\phi_{n}(\omega)}\,\:, \label{eq:TFbackaction}
\end{equation} 
and the resulting displacement can  be
considered as the coherent sum of the contributions of independent harmonic 
oscillators, each corresponding to a normal mode of the system and moving 
under the same driving force $F_{BA}(t)$. We note that the displacements contributed by different modes could also be in opposite directions, but it is only the resulting sum that is  physically observable.

\subsection{Readout optimization and SQL}
\label{subsec:readoutoptimization}
 The readout noise properties are defined in terms of the single sided power spectral densities $S_{xx}(\omega)$ and  $S_{ff}(\omega)$. Here $S_{xx}(\omega)$ is the equivalent input displacement noise power spectral density contributed by the readout; $S_{ff}(\omega)$ is the equivalent force noise power spectrum density due to the
readout that drives the system through the transfer function  $T_{BA}$ [Eq. (\ref{eq:TFbackaction})]. Under the hypothesis of  uncorrelated noise sources, the readout contribution to the noise on the observable $X$ is:
\begin{equation}
\label{eq:ampnoise}
S_{XX}(\omega) =S_{xx}(\omega)+|T_{BA}(\omega)|^2 S_{ff}(\omega)\,.
\end{equation}

A useful figure for expressing the readout performance is its energy resolution expressed as  number of energy quanta:
\begin{equation}
\label{eq:ampnoiseres}
\epsilon_r (\omega)= \frac{\sqrt{S_{xx}(\omega) S_{ff}(\omega)}}{\hbar}\,.
\end{equation}
The uncertainty relation for a continuous linear measurement requires that \cite{braginsky}:
\begin{equation}
\epsilon_r (\omega) \,\geq 1 \,,
\end{equation}
and the readout is called ``quantum limited'' when this limit is achieved.
It is also helpful to define the readout noise stiffness $\kappa_r$ as:
\begin{equation}
\label{eq:ampnoiseimp}
\kappa_r (\omega) = \sqrt{S_{ff}(\omega)/ S_{xx}(\omega)}\,.
\end{equation}
If we now write the  power spectral densities $S_{xx}(\omega)$ and  $S_{ff}(\omega)$ in terms of  $\epsilon_r$ and  $\kappa_r$:
\begin{eqnarray}
\label{eq:inversenoise}
 S_{xx}(\omega)&=&\hbar \,\epsilon_r (\omega) / \kappa_r(\omega)\\
S_{ff}(\omega)&=&\hbar \, \epsilon_r (\omega)\,\kappa_r (\omega)\,,
\end{eqnarray}
the Eq. (\ref{eq:ampnoise}) becomes:
\begin{equation}
\label{eq:ampnoise1}
S_{XX}(\omega) =\hbar\, \epsilon_r (\omega) \,\bigg[\frac{1}{\kappa_r(\omega)}+ {|T_{BA}(\omega)|^2}{\kappa_r(\omega)}\bigg]\,.
\end{equation}
It is straightforward to demonstrate that for each  $\omega$ the readout noise contribution is minimized, giving the SQL  \cite{braginsky} for the detector, if the following equations are both satisfied:
\begin{eqnarray}
\label{eq:SXXmin_epsilon}
\epsilon_r (\omega_{})&=&1\\
\label{eq:SXXmin_kappa}
\kappa_r (\omega_{})&=&\frac{1}{|\,T_{BA}(\omega_{})|}\,.
\end{eqnarray}
Equivalently the following must hold:
\begin{eqnarray}
\label{eq:Sxxsqlnarrow}
 S_{xx}(\omega_{})&=&\hbar\,|\,T_{BA}(\omega_{})|\nonumber\\
 S_{ff}(\omega_{})&=&\frac{\hbar}{|\,T_{BA}(\omega_{b})|}
\end{eqnarray}
These relations show that, as obvious, the best detector performances require the use of  a quantum limited readout and, less trivial, the noise stiffness must be properly matched to the test mass mechanical impedance $T_{BA}(\omega_{})$.  The total displacement noise becomes:
\begin{equation}
\label{eq:sxxSQL}
 S_{XX} (\omega_{})
=2\,\hbar\,|\,T_{BA}(\omega_{})|\,,
\end{equation}
and can be transformed into an equivalent input GW  spectral density through the GW transfer function $H_{GW}(\omega)$ [Eq. (\ref{gwtransfer})]. The detector sensitivity is then:
\begin{equation}
\label{eq:shhSQL}
 S_{hh} (\omega_{})
=2\,\hbar\,\frac{|\,T_{BA}(\omega_{})|}{|\,H_{GW}(\omega_{})|\,^2}.
\end{equation}

We point out that the SQL here derived is calculated taking into account the volume properties of test-mass, signal and readout: this is in contrast to the usual derivation of the SQL for a  GW detector, in which a 1-dimensional model is considered.

In principle a wide bandwidth detector, designed to be sensitive in a angular frequency range $\omega_{min}\div \omega_{max}$ could reach the SQL at every frequency, provided that a readout satisfying  Eq. (\ref{eq:Sxxsqlnarrow}) on the full interval $\omega_{min}\div \omega_{max}$ can be designed. The elastic body transfer function $T_{BA}(\omega)$  is a rapidly varying function (as we show below in Fig.  \ref{fig:transferfunctions}b): sharp magnitude peaks, corresponding to the resonant normal modes of the system, alternate with  magnitude dips called antiresonances. Unfortunately the noise properties of a practical wide bandwidth readout have a smooth frequency dependence over the detector sensitivity range and cannot match these structures of the transfer function, preventing the possibility of reaching the SQL on a wide frequency range. 

The origin of antiresonances is well explained within the modal expansion model. We show in Appendix \ref{appendix:modal} that in the back action response $T_{BA}(\omega)$ there must be an antiresonance following a resonance, without exception and regardless of the complexity of structures.  An optimized readout should have a high back action force noise spectral density  at the anti-resonances frequencies, where the detector is quite insensitive. On the contrary it must have low back action noise at frequencies where resonances of $T_{BA}(\omega)$ occur, in order to avoid the resonance amplification of the readout force noise.  The optimization process will approach the best performances by averaging among these conflicting requirements.

The frequency dependence of the readout noise properties depends on its actual design. In order to show the optimization strategy, we work out here the simple case of a readout with constant noise properties:
\begin{eqnarray}
\epsilon_r(\omega)&=&\epsilon_{0} \\
\kappa_r(\omega)&=&\kappa_{0} \,.
\end{eqnarray}

The readout contribution to the  equivalent input displacement noise on the variable $X$ becomes now:
\begin{equation}
\label{eq:ampnoisewideband}
S_{XX}(\omega) =\hbar\, \epsilon_0   \,\bigg[\frac{1}{\kappa_0}+ {|T_{BA}(\omega)|^2}{\kappa_0}\bigg]\,,
\end{equation}
and the detector sensitivity is simply:
\begin{equation}
\label{eq:shhwideband}
S_{hh}(\omega) =\frac{\hbar\, \epsilon_0 } {|H_{GW}(\omega)|^2 }\,\bigg[\frac{1}{\kappa_0}+ {|T_{BA}(\omega)|^2}{\kappa_0}\bigg]\,.
\end{equation}
We find again that for best performances the use of a quantum limited readout $(\epsilon_0=1)$ is required, and the noise stiffness $\kappa_0 $ must be properly matched to the readout. The optimization strategy necessarily depends on the characteristic of the expected GW sources. In fact for a given GW signal $\widetilde{h}(\omega)$, the measurements signal to noise ratio $S/N$ is defined as:
\begin{equation}
\label{eq:snr}
\frac{S}{N}=\sqrt{\frac{2}{\pi}\,\int_{0}^{+\infty} \, \frac{|\widetilde{h}(\omega)|^2}{S_{hh}(\omega)}\,d\omega } \,.
\end{equation}

\subsection{Thermal noise}
In order to reach the sensitivity shown in Eq. \eqref{eq:shhwideband}  technical noises must be reduced down to a negligible level. Among all possible technical noise sources, the one due to the fluctuations of the body volume (the so-called brownian or thermal noise) often ends up as a barrier preventing the achievement of the SQL.

In the case of a system  at equilibrium with a bath at temperature T, with a single coordinate $p$ and a force $F$ that represents the interaction between the system and the externals, the single sided thermal noise power spectrum  of the coordinate is predicted by the fluctuation-dissipation theorem \cite{callen51} to be:
\begin{equation}
\label{thermalpowerdef}
 S_{pp}(\omega)=-\frac{4\,k_B \,T}{\omega}\;\Im m[T_F(\omega)].
\end{equation}
where $T_F(\omega)$ is the transfer function:
\begin{equation}
T_F(\omega)= \frac{\widetilde{p}(\omega)}{\widetilde{F}(\omega)}.
\end{equation}

When the system coordinate $X$ is a linear combination weighted by a function $\textbf{P}(\textbf{r})$, as in Eq. (\ref{observabledef}), the thermal noise on $X$ can be evaluated by the system response to a force $F(t)\, {\textbf{P}(\textbf{r})}$ applied to the test mass surface \cite{levin}. The thermal noise power spectral density on the output variable is then given by the transfer function $T_{BA}(\omega)$ [Eq. (\ref{eq:TFbackaction})]:
\begin{eqnarray}\label{thermalpower}
&&S_{XX}(\omega, T)= -\frac{4\,k_B \,T}{\omega}\;\Im m[T_{BA}(\omega)]\,\\
&=&-\,\frac{4\,k_B \,T}{\omega }\,\sum_{n}\,\frac{-\,\omega_{n}^2\,
\phi_{n}(\omega)}{M} \,\frac{ |\int_s\textit{d}\,
s\,
{\textbf{P}(\textbf{r})}\,\cdot
\,\textbf{w}_n(\textbf{r}) \,|^2}{(\omega^2_{n}-\omega^2)^2+ \omega^4_{n}\,
\phi_{n}^2(\omega)}\,. \nonumber 
\end{eqnarray}

The mechanical thermal noise of the system can be reduced in principle below the readout contributed noise by reducing the detector temperature $T$. If the thermal noise evaluated in Eq. (\ref{thermalpower}) must be well below the  readout SQL noise given by Eq. (\ref{eq:sxxSQL}), then:
\begin{equation}
\label{eq:Tcondition}
-\frac{4\,k_B \,T}{\omega}\;\Im m[T_{BA}(\omega)]\,\ll \,2\,\hbar\,|T_{BA}(\omega_{})|\,.
\end{equation}

Notice that the transfer function $T_{BA}$ enters in both sides of Eq. (\ref{eq:Tcondition}), but with different meanings: the left-hand side is essentially the incoherent superposition of the thermal noises generated by each normal mode of the system, while the right-hand side is the coherent superposition of the normal mode responses. 

The effect of the detector thermal noise on the $S/N$ can be easily evaluated if the sensitivity $S_{hh}$ [Eq. (\ref{eq:shhwideband})] is generalized as:
\begin{equation}
\label{eq:shhT}
S_{hh}(\omega,T) = 
\frac{\hbar\, \epsilon_0 \,\bigg[\frac{1}{\kappa_0}+ {|T_{BA}(\omega)|^2}{\kappa_0}\bigg] + S_{XX}(\omega,T)} {|H_{GW}(\omega)|^2 }\,, 
\end{equation}
where the detector thermal noise $S_{XX}(\omega,T)$ is given by Eq. (\ref{thermalpower}).
In the case of structural damping the thermal noise contribution to $S_{hh}(\omega,T)$ is proportional to the product $T \phi$.

\section{A case study: the single-mass DUAL detector}
\label{section:singledual}

\begin{figure}[t!]
\includegraphics[width=8.6cm,height=4.1cm]{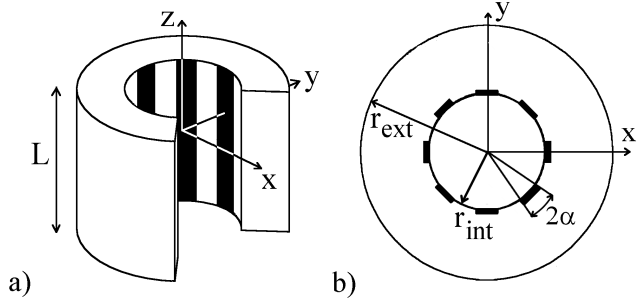}
\caption{\label{fig:cylinder}  
(a) 
Test mass and readout: the displacements induced by a GW propagating along the $z$ axis are measured in 8 regions (in black) spanning the whole cylinder height $L$. 
(b) 
Section of the detector: the angle $\alpha$ determines the extent of the measuring surfaces and represents a compromise between the requirements of wide sampling area and the possibility of implementing 8 independent sampling areas on the test mass. The value $\alpha =0.2$ rad is used throughout the paper.}  
\end{figure}

On the basis of the optimization process described in Section \ref{subsec:readoutoptimization}, we can estimate the  SQL sensitivity of  specific detector configurations, provided that the eigenvalue problem Eq. (\ref{eigenproblem0}) for the test mass is solved. As an example we consider an hollow cylinder test mass (Fig. \ref{fig:cylinder}a). If we limit our evaluation to GW propagating along the $z$ axis, the symmetry axis of the system, the corresponding force does not depend on \textit{z}: the system response can be well described by plane strain solutions, where the displacements are functions of \textit{x} and \textit{y} only and the displacement along \textit{z} vanishes. The analytical plain strain solutions  are well known \cite{gazis} and we use them to evaluate the detector sensitivity; we make use of FEM simulations to confirm  that the main features of the detector are maintained when a real 3-dimensional test mass is considered.

We show in the following that, thanks to the effect of its  lowest order quadrupolar resonant modes, this \emph{single} test mass detector offers the same advantages of the two nested mass configuration (namely the Dual detector), recently proposed \cite {cerdonio1,bonaldi1,briant}: the large bandwidth, set by the frequency  difference between the lowest order GW sensitive modes of each test mass, and the back action reduction effect, determined by the antiresonance frequency placed between these modes.  
To fully exploit the DUAL detector properties we consider to implement a pair of selective readouts. These readouts \cite{bonaldi1} are sensitive to the two independent components of the quadrupolar modes and reject other classes of modes, which are not excited by GWs. 

\begin{figure}[t!]
\includegraphics[width=8.6cm,height=4.7cm]{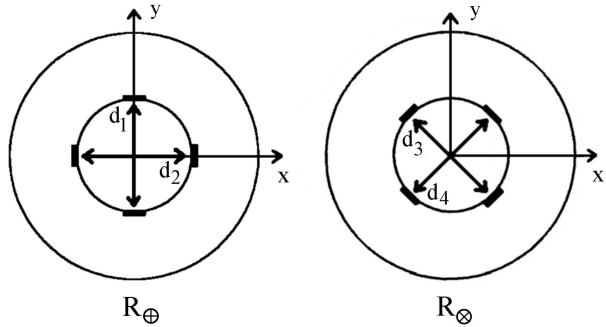}
\caption{\label{fig:readouts}  
Sections of the detector showing the detail of the readout measurement strategy. The measuring surfaces are arranged in two independent readout configurations: the readout $R_\oplus$ measures the difference $X_\oplus=d_2-d_1$, while $R_\otimes$ measures the difference $X_\otimes=d_4-d_3$. It will be shown that $R_\oplus$ and $R_\otimes$ are sensitive respectively to the $+$ and $\times$ components of a GW propagating along the $z$-axis. Both readouts are mainly sensitive to the quadrupolar modes of the test mass.
}
\end{figure}

The selective readouts $R_\oplus$ and $R_\otimes $(Fig. \ref{fig:readouts}) measures the inner diameters of the hollow cylinder  and combine them to obtain $X_\oplus=d_2-d_1$ and $X_\otimes=d_4-d_3$. In each diameter evaluation the test mass inner surface displacement is averaged over two opposite areas spanning the full cylinder length. These detection schemes are implemented by the weight function:
\begin{eqnarray}
\label{eq:Pweightplus}
\mathbf{P}_\oplus(\mathbf{r})&=& \,\delta(r-r_{int})\,P_{\theta}(\theta)\,\mathbf{i}_r \\
\label{eq:Pweightcross}
\mathbf{P}_\otimes(\mathbf{r})&=& \,\delta(r-r_{int})\,P_{\theta } (\theta + \frac{\pi}{4})\,\mathbf{i}_r\,,
\end{eqnarray}
with:
\begin{equation}
P_{\theta}=\frac{1}{S_0 }\sum_{m=0}^1\, \sum_{n=0}^4 (-1)^{n+m}
\,\Theta\,[\,\theta+(-1)^{m}\alpha -n\frac{\pi}{2}\, ] \,,
\label{eq:Pangweight}
\end{equation}
where $\Theta (x)$ represents the unit step function and $2\alpha$ is the angle subtended by the each measured surfaces (Fig. \ref{fig:cylinder}b). The value  $\alpha =0.2$ rad is used in the following:  it represents a good compromise between the requirements of wide sampling area and the possibility of implementing two readouts rotated by $\pi/4$ on the same test mass.  The normalization $S_0$ is the area of each measured surface, evaluated as $S_0\simeq 2 \alpha \,r_{int}\,L$; as shown in Figure \ref{fig:cylinder}, $r_{int}$ is the inner radius of the hollow cylinder. It is straightforward to demonstrate that, according to Eq. (\ref{eq:shhT}), the detector sensitivity does not depend on the chosen normalization $S_0$.  
 
\subsection{Detector response}
\label{sec:planestrain}
The plain strain normal modes of the motion of the cylinder in the $x-y$ plane  are solution of the eigenstate problem [Eq. (\ref{eigenproblem0})], when the following additional boundary conditions are applied:
\begin{eqnarray}
\label{eq:constraints}
 u_z \quad\quad &\equiv& 0 \nonumber \\
 \frac{\partial(u_r,u_z,u_\theta)}{\partial z}\,&\equiv &\,0\:,
\end{eqnarray}
where $(u_r,u_\theta,u_z)$ are components of the displacement $\textbf{u}(\textbf{r})$ in a cylindrical coordinate system.
The analytical expressions of the normal modes of an hollow cylinder in the plane strain approximation are functions of the kind \cite{gazis}:
\begin{eqnarray}
\label{twinmodesx}
\textbf{w}^\oplus_{a,\,n}(\textbf{r})&=&f_{a,\,n}(r)\,\cos\,(a\,\theta)\:
{\textbf{i}}_r \: + \:g \,_{a,\,n}(r)
\,\sin \,( a \,\theta )\:\textbf{i}_\theta \label{twinmodes+} \\
\textbf{w}^\otimes_{a,\,n}(\textbf{r})&=&-f_{a,\,n}(r)\,\sin\,(a\,\theta)\,
{\textbf{i}}_r\: +\:g\,_{a,\,n} (r) \cos  \,(a
\,\theta)\,\textbf{i}_\theta\,, \nonumber 
\end{eqnarray}
where the functions $f_{a,\,n}, g\,_{a,\,n}$ are linear combinations of Bessel functions of the coordinate $r$, with coefficients determined by the boundary conditions. The integer $a$ represents the angular symmetry of the mode, while $n$ identifies the mode order within the angular family; modes within a given family are ordered by increasing frequency with increasing $n$. The orthogonal displacement fields $\textbf{w}^\oplus_{a,\,n},\textbf{w}^{\otimes}_{a,\,n}$  represent the same radial distribution of the deformation, mutually rotated by $\frac{\pi}{2a}$: for this reason they share the same eigenvalue $\omega_{a,\,n}$.  We call $\nu_{a,\,n}=\omega_{a,\,n}/(2\pi)$ the resonant frequency of the mode.  The displacement $\textbf{w}^s_{a,\,n}$ is an element of an orthogonal complete system, and is normalized to satisfy the condition  (with $s,\,t=\oplus$ or $\otimes$):
\begin{equation}
\int_V \textit{d}V \, \rho\;
\textbf{w}^s_{a,\,n}(\textbf{r})\cdot\textbf{w}^t_{b,\,m}(\textbf{r})\,
=M \,\delta_{s\,t}\,\delta_{a\,b}\,\delta_{n\,m}\:.
\label{ortho}
\end{equation}
Any plane strain displacement $\textbf{u}_{p}$ may be written as
linear superposition of these basis functions, with time dependent
coefficients determined by the force acting on the body:
\begin{equation}
 \textbf{u}_{p}(\textbf{r},t)= \sum_{s,\,a,\,n} \textbf{w}^s_{a,\,n}(\textbf{r})\,q^s_{a,\,n}(t)\, .
 \label{sumsolutionplane}
\end{equation}
The plain strain approximation holds in presence of an external force when this vanishes, with its first derivatives, along the $z-$axis:
\begin{equation}
\label{eq:force_z_condition}
\textbf{G}_r(\textbf{r})\cdot\,\textbf{i}_z=0 \quad  \quad 
\frac{\partial\,\textbf{G}_r(\textbf{r})}{\partial\,z}=0 \,.
\end{equation}
Moreover the same conditions must hold for the weight function of the readout used to measure the displacement:
\begin{equation}
\label{eq:readout_z_condition}
\textbf{P}(\textbf{r})\cdot\,\textbf{i}_z=0 \quad  \quad
\frac{\partial\,\textbf{P}(\textbf{r})}{\partial\,z}=0\,\,.
\end{equation}
The plane strain solutions are exact for an infinite length cylinder, as the fulfillment of   plane strain condition requires the application,  over the terminal sections, of tension or pressure adjusted so as to keep constant the length of all longitudinal filaments (that is the body height). In the case of a finite length cylinder the solutions can be considered exact only if a surface force opposite to the internal stress is applied to the end surfaces of the cylinder. According to the Saint-Venant principle \cite{Love}, we infer that the plane strain approximation is good when   the length of the cylinder is large compared to its diameter. In this case   stress and strain  in the interior are practically independent of the presence of a force distribution applied on the ends, in all the portions of the cylinder except comparatively small slices near its ends.

\begin{figure}[ht!]
\includegraphics[width=8.6cm,height=4.3cm]{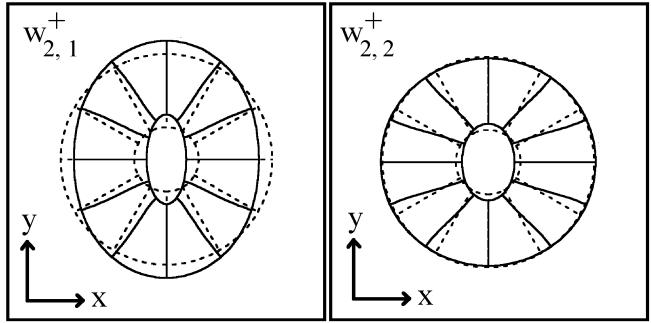}
\caption{\label{fig:duemodi}  
Modal shape of the first two plane strain quadrupolar modes. The continuous lines show the deformation of internal and external surfaces and of 12 radial lines traced ideally inside the cylinder section. The shape of the cross section at null deformation is  shown as a dashed line.  Both modes induce deformations of the inner surface of the cylinder and then  contribute  to the detector output.  (a) First quadrupolar mode $\textbf{w}^\oplus_{2,\,1}(\textbf{r})$: both the inner and outer surfaces change their shape.  The cylinder thickness $r_{ext}-r_{int}$  remains approximatively constant (flexural mode). (b) Second quadrupolar mode $\textbf{w}^\oplus_{2,\,2}(\textbf{r})$:  the outer surface remains essentially unchanged and the deformation involves only inner portions of the cross section. This deformation is then obtained by a change of the cylinder thickness (bulk mode).}
\end{figure}

On the basis of the solutions Eq. (\ref{twinmodes+}), for instance, we can evaluate the effect of a  $+$ polarized GW propagating along the $z$-axis, say $\textbf{W}^+(\textbf{r})$, given by Eq. (\ref{gwaveplus}) with $\psi =0$. 
This wave can excite only the family of the $\mathbf{w}^\oplus$ plane strain quadrupolar ($a=2$) modes, as we have: 
\begin{equation}
\label{eq:gwplusconvolution}
 \int_V \textit{d}V \, \textbf{W}^+(\textbf{r})\cdot\textbf{w}^\otimes_{2,\,n}(\textbf{r})=0 \,,
\end{equation}
due to the quadrupolar symmetry of the force field. The resulting displacement would be the superposition of the displacement contributed by  an infinite number of  modes with $a=2$, but we find that the main contribution is given by the two  modes of lowest frequency, $\textbf{w}^\oplus_{2,\,1}(\textbf{r})$ and $\textbf{w}^\oplus_{2,\,2}(\textbf{r})$. These modes induce an elliptical strain of the inner surface of the cylinder, as shown in Fig. \ref{fig:duemodi}. We remind that these modal shapes remain unchanged over the whole frequency range (namely also far from their resonance),  whatever the frequency of the driving force, while their relative strength varies according to Eq. (\ref{eq:qsolution}). 

As discussed in Appendix \ref{appendix:modal}, the relative phase between the two contributions depends on the sign of the modal constants, which in this case, [according to Eq. (\ref{gwtransfer})] are determined by the convolution of the GW spatial force $\textbf{W}(\textbf{r})$  over the modal shape \textit{and} by the readout weight function $\textbf{P}_\oplus(\textbf{r})$ [Eq. (\ref{eq:Pweightplus})]:
\begin{equation}
\label{eq:modalgwconstants}
C^\oplus_{2,\,n}\propto \big[\int_V \textit{d}V \, \textbf{W}(\textbf{r})\cdot\textbf{w}^\oplus_{2,\,n}(\textbf{r})
\big]\; \big[\int_S\textit{d}s\, {\textbf{P}_\oplus(\textbf{r})\,\cdot
\,\textbf{w}^\oplus_{2,\,n}(\textbf{r})}  \big] \,.
\end{equation}
The two modes induce a similar deformation of the inner surface, but given by  different deformations of the cylinder interior: the $\textbf{w}^\oplus_{2,\,1}$ is a flexural mode and it does not change the cylinder thickness $r_{ext}-r_{int}$, while $\textbf{w}^\oplus_{2,\,2}$ is a bulk mode which changes the thickness.   
It is then not surprising that the analytical evaluation of Eq. (\ref{eq:modalgwconstants}) results in two modal constants with \textit{opposite sign}. For this reason, when the detector is driven by a GW at frequency $\nu_{gw}$  we recover the characteristics of a Dual detector \cite{cerdonio1}; namely the contribution  of the two modes:
\begin{enumerate}
\item[-] subtracts at low frequencies ($\nu_{gw}<\nu_{\,2,\,1}$), when they  follow in phase the driving force. 
\item[-] is enhanced when the modes  get out of phase in the interval $(\nu_{{}\,2,\,1} \div \nu_{\,2,\,2})$, thanks to the $\pi$ phase lag of the mode $\nu_{\,2,\,1}$  which is driven above its resonant frequency. This frequency range determines approximatively the detector bandwidth.  
\item[-] subtracts at higher frequencies ($\nu_{gw}>\nu_{\,2,\,2}$) when both modes get out of phase because they are driven above their resonance. 
\end{enumerate}

\begin{figure}[ht!]
\includegraphics[width=8.6cm,height=8.6cm]{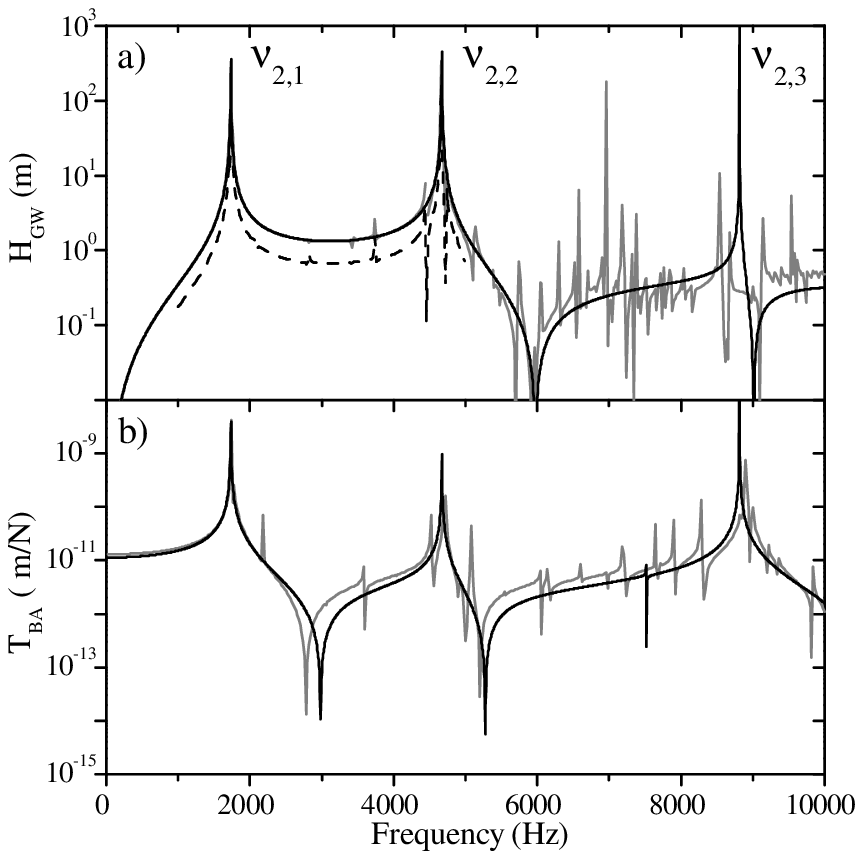}
\caption{\label{fig:transferfunctions} Transfer functions for the $R_\oplus$ readout with $\alpha=0.2\;$rad describing the dynamical response of an hollow cylinder single-mass DUAL detector (material: molybdenum,  outer radius $r_{ext}=0.5$ m, heigth $L=3$ m,  form ratio $r_{int}/r_{ext}=0.3$). (a) Analytical (continuous line) and FEM evaluated (gray line) GW transfer function for a signal propagating along z-axis. Dashed line:  3D FEM evaluated response under a $+$ polarized GW traveling along the $y$-axis. (b) Back action transfer function. Continuous line: analytical plain strain evaluation. Gray line:  3D FEM simulation.
}
\end{figure}

The resulting GW transfer function is shown in Figure \ref{fig:transferfunctions}a: the sensitivity peaks due to the quadrupolar modes  can be seen clearly.  Figure \ref{fig:transferfunctions}b   shows the back action transfer function $T_{BA}$, evaluated on the basis of the readout weight function $\textbf{P}_\oplus(\textbf{r})$ with $\alpha=0.2$ rad.  An antiresonance shows up within the range $\nu_{2,\,1}\div \nu_{2,\,2}$, as expected on the basis of the discussion in Appendix \ref{appendix:modal}.   The frequency of the antiresonance  depends not only on the modal constants of the first two quadrupolar modes, but also by the contribution of all modes sensed by the readout. We observed that a convergence better than 1\% is obtained if all modes with frequency below 30 $kHz$ are considered in the evaluation (about 15 modes).

In Figure \ref{fig:transferfunctions} we also show the transfer functions obtained by a 3 dimensional FEM harmonic analysis. These harmonic analysis are performed in the range $0\div 10\;$kHz using the ANSYS \cite{ansys} \textit{full method} of solving dynamic equations, proved to be an accurate method of evaluating the elastic body dynamical impedance \cite{yamamoto}.  The FEM code calculates the transfer functions from the displacement averaged over the readout surface according to the weight function Eq. (\ref{eq:Pweightplus}). To evaluate $T_{BA}$, a unit external force is applied onto the readout surfaces according to Eq. (\ref{eq:backaction}). The GW volume force field Eq. (\ref{gwaveplus}) is   applied to evaluate $H_{GW}$. As expected, the resonant modes found in the plane strain approximation are essentially confirmed by the 3D FEM evaluation, but other   modes appear that cannot be obtained in the plane strain framework. Thanks to the readout capability of rejecting non
GW sensitive modes, the displacement induced by these additional modes is efficiently averaged out and the detector dynamical performances do not differ from that evaluated in the plain strain approximation. This fact confirms that the body dynamics is modified only weakly by the absence of the balancing force applied at the bottom and top surfaces of the cylinder.

The FEM code is not limited to plane strain evaluation and can also calculate the  sensitivity pattern of the detector for GWs arriving  from an arbitrary direction $(\theta,\, \varphi)$ with an arbitrary polarization $\psi$. 
Following the approach introduced in ref. \cite{forward}, we separate the radiation into its two linear polarization components, described by $\psi=0$ and $\psi=\pi /4$ in the wavefront plane of the GW. When $(\theta,\,\varphi)=(0,\,0)$  these force fields reduce respectively to the $+$ and $\times$ GW  described by Eq. (\ref{gwaveplus}).

\begin{figure}[ht!]
\includegraphics[width=8.6cm,height=6.5cm]{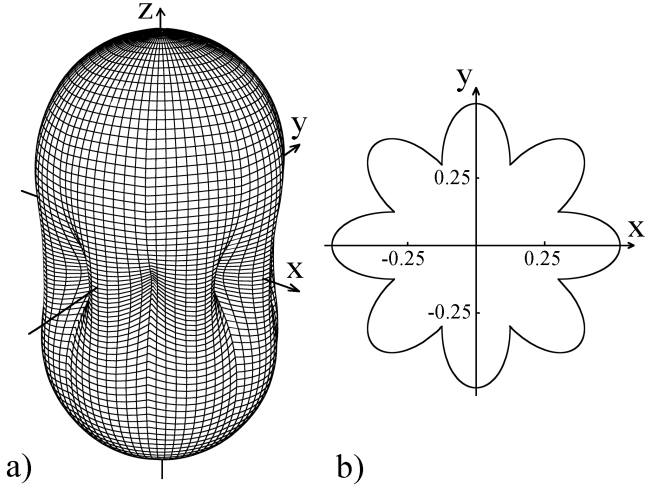}
\caption{\label{fig:antennapattern} 
Antenna pattern  of an hollow cylinder single-mass Dual detector (material molybdenum, outer radius $r_{ext}=0.5$ m, heigth $L$ = 3 m, form ratio $r_{int}/r_{ext}=0.3$, $\alpha$=0.2rad)
 (a) Antenna pattern of the detector equipped with the readouts $R_\oplus$ and $R_{\otimes}$. The antenna pattern is represented by the modulus of the vector to the plotted surface. Per each direction, we show the maximum of the antenna patterns of the readout channels for circularly polarized signals. 
 (b) Section of the antenna pattern across the $x-y$ plane. This is the worst case for the detector, which is maximally sensitive for GW traveling along the $z$-axis; nonetheless  no blind directions shows up. The response  ranges from 30\% to 55\% of the optimal case.
}
\end{figure}

The frequency response to a GW is scaled down for a wave from any direction with respect to the z-propagating case, as shown in Figure \ref{fig:transferfunctions}a  in the case of a wave traveling along the $y$-axis.  We evaluated the scaling factor as a function of the signal direction $(\theta,\,\phi)$ and polarization $\psi$ at a fixed frequency (3000 Hz). For each readout $R_\oplus$ and $R_\otimes$, the normalized output for a circularly polarized GW is obtained by quadratically averaging the response to $\psi=0$ and $\psi=\pi /4$ waves. The antenna pattern shown in Figure \ref{fig:antennapattern}a refers to circularly polarized signals and shows the maximum response between the two readouts. 

The Dual detector is then omnidirectional,  in the sense that no blind directions show up in its antenna pattern. On the other hand it is not isotropic, and the lowest response for a circularly polarized GW is still 30\% of the response to an optimally oriented GW, i.e. propagating along z. In general, the two linear polarization components of the GW excite two independent linear combinations of $\textbf{w}^\oplus_{2,\,n}(\textbf{r})$ and $\textbf{w}^\otimes_{2,\,n}(\textbf{r})$, which are separately read by $R_\oplus$ and $R_\otimes$ respectively. Therefore, it is relevant to point out that the single-mass Dual detector can resolve both amplitude polarizations of the impinging GW if its propagation direction $(\theta, \phi)$ is known. This is not the case of an interferometric or resonant bar detector, which measures only one linear polarization component of the GW.

The presence of two independent readout channels which monitor different signal polarizations also enhance the possibilities of setting up vetoing strategies in joint observations with other GW detectors, by exploiting Riemann tensor properties to discriminate noise events. In addition, the two independent readouts are affected by independent intrinsic noise sources, such as thermal and amplifier noises, so that they can be considered as coming from different detectors as long as environmental disturbances are negligible.

\subsection{Detector design and optimization}
\label{subsec:optimization}
\begin{figure}[t!]
\includegraphics[width=8.6cm,height=5cm]{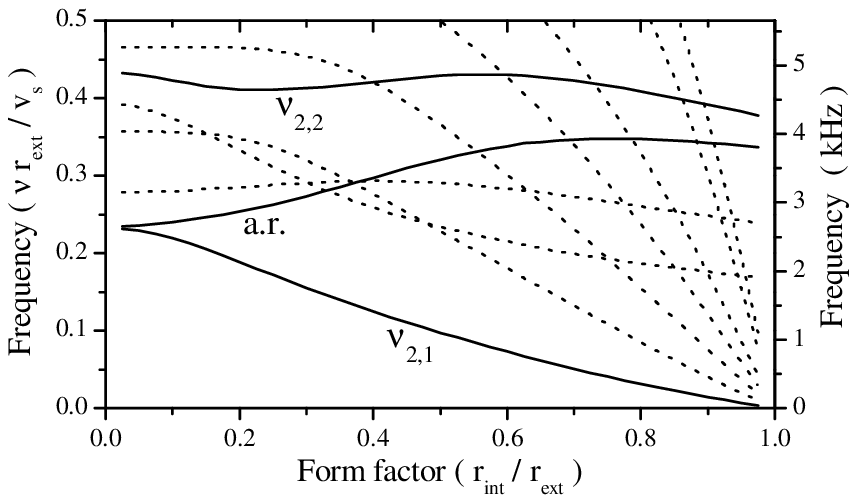}
\caption{\label{fig:modechart}  
Frequency of the normal modes in the plane strain approximation. The left axis shows the reduced value of the frequency, while the right axis shows the actual value of the frequency in the case of a molybdenum  test mass with $r_{ext}$=0.5 m and $v_s=5650$ m/s.  The frequency $\nu_{\,2,\,1}$ of the first quadrupolar mode (flexural mode) is strongly dependent on the form factor, while the frequency $\nu_{\,2,\,2}$ of the second quadrupolar mode (bulk mode) is essentially constant. The antiresonance frequency (a.r.) is evaluated accounting for the contribution of the number of modes needed to obtain a 5\% convergence. We point out that while the mode frequencies are solutions of the eigenvalue problem Eq. (\ref{eigenproblem0}) and do not depend on the chosen readout, the antiresonance frequency is evaluated specifically for the selective readout shown in Figure \ref{fig:cylinder}b. A number of non GW sensitive modes are also shown as dotted lines.
}
\end{figure}

In this section we determine the optimal detector figures on the basis of the analysis developed in Section \ref{section:SQLmodel}.  As usual a low dissipation material is required to reduce the effect of the thermal noise. Molybdenum represents an interesting choice, as it shows high cross-section for GWs and low mechanical losses at low temperatures: ref. \cite{duffyMo} reports a value $T\phi  < 5 \times 10^{-9}\;$K for acoustic modes down to 50 mK.  

In this section we choose to optimize the detector geometry by requiring a specific frequency band, namely the interval 2-5 kHz. This requirement fixes the frequency of the first two quadrupolar modes as $\nu_{\,2,\,1}\simeq$ 2 kHz and $\nu_{\,2,\,2}\simeq$ 5 kHz.
As shown in Figure \ref{fig:modechart}, the frequency $\nu_{\,2,\,2}$  depends essentially on the outer radius $r_{ext}$ and on the material sound velocity $v_s$. A frequency $\nu_{\,2,\,2}\sim 5$ kHz could be obtained with a  molybdenum test mass with $r_{ext}=0.5$ m.
The height of the cylindrical test mass is fixed to $L=3$ m. The inner radius, through the form factor $r_{int}/r_{ext}$, determines the frequency of the first quadrupolar mode and the placement of the antiresonance within of the bandwidth.  In order to better understand the effect of the detector geometry on the sensitivity, we  evaluate the sensitivity of the  detector  for  different values of the form factor in the range $0.1<r_{in}/r_{ext}<0.8$.

\begin{figure}[ht!]
\includegraphics[width=8.6cm,height=13.1cm]{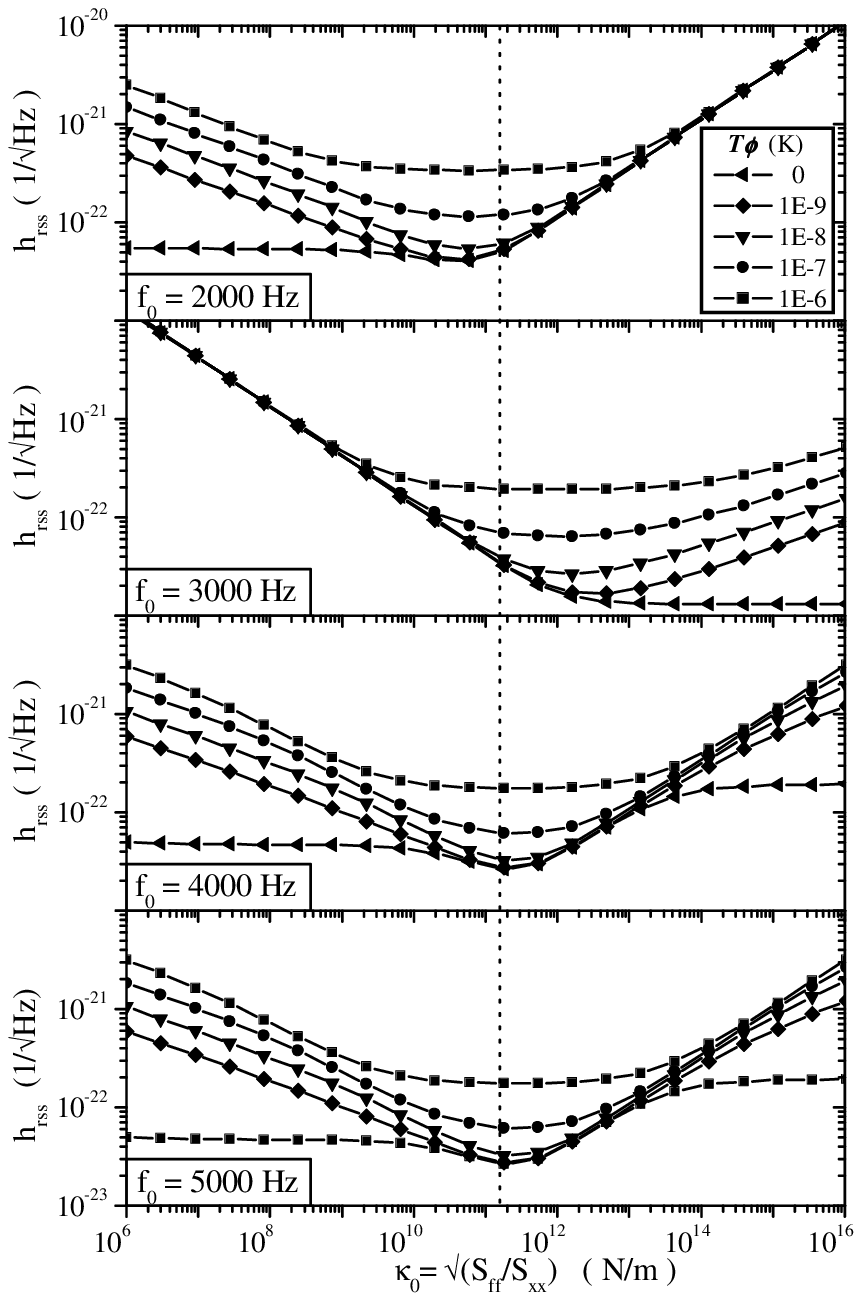}
\caption{\label{fig:ottimizzazione}  
Minimum detectable (at SNR=1) root sum square amplitude $h_{rss}$ of a sine-Gaussian test waveform, impinging the detector at 4 different values of the central frequency $f_0$.   The calculation was performed for a detector form factor $r_{int}/r_{ext}=\;0.3$, equipped with a quantum limited readout $(\epsilon_0=1)$. The value $\kappa_0\simeq 1.7\times 10^{11}$ N/m, indicated by the dotted vertical line, gives the best $h_{rss}$ value averaged over the 4 test frequencies. This optimal value of the noise stiffness is independent of the $T\phi$ value and corresponds to a readout displacement power noise of $S_{xx}\simeq 6\times 10^{-46}$  m$^2$/Hz and back action force  power noise of $S_{ff}\simeq 1.8 \times 10^{-23}\;$ N$^2$/Hz. 
}
\end{figure}

As discussed in Section \ref{subsec:readoutoptimization}, the  optimization process requires a balancing of the noise figures.   For this sake we must evaluate as a function of $\kappa_0$ the minimum detectable value of a specific GW signal by the use of Eq. (\ref{eq:snr}). As in the 2-5 kHz range burst signals are expected from compact stellar object, we consider a sine-Gaussian as test waveform:
\begin{equation}
\label{eq:sinegaussian}
h(t)=h_0 \,sin[\omega_0 \,(t-t_0)]\,e^{-\frac{(t-t_0)^2}{\tau^2}}\,,
\end{equation}
with central frequency $f_0=\omega_0/2\pi$ and $\tau=2/f_0$. For GW detectors it is customary to quote performance in terms of the  minimum detectable value (at S/N=1) of the root sum square amplitude $h_{rss}$:
\begin{equation}
\label{eq:hrss}
h_{rss}=\sqrt{\int _{-\infty}^{+\infty} dt \, |h(t)|^2}=\sqrt{\frac{1}{\pi}\int _{0}^{+\infty} d\omega \, |h(\omega)|^2}\,.
\end{equation}
In order to estimate the detector performances on its full bandwidth,  4 different central frequencies were used for the burst of Eq. (\ref{eq:sinegaussian}). The calculation was also performed  as a function of the amount of dissipation ($T\phi$ value) considered in the test mass.  As shown in Figure \ref{fig:ottimizzazione} for the case $r_{int}/r_{ext}=\;0.3$ and $\epsilon_0=1$, the optimal value of the noise stiffness is $\kappa_0\simeq 1.7\times 10^{11}$ N/m. This number is  independent of the $T\phi$ value and, according to Eq. (\ref{eq:inversenoise}), corresponds to a readout displacement power noise of $S_{xx}\simeq 6\times 10^{-46}$  m$^2$/Hz and  back action force power noise of $S_{ff}\simeq 1.8 \times 10^{-23}$ N$^2$/Hz. 

We see in Figure \ref{fig:ottimizzazione} that, at the optimal $\kappa_0$,  the thermal noise induced by a finite dissipation value $T\phi = 10^{-8}\;$K  reduces the sensitivity of less than 20\%. This dissipation is a good compromise between sensitivity and feasibility, and can be assumed as the reference $T\phi$ value for the case $\epsilon_0=1$. In the general case the optimal $T\phi$  value is determined by the readout energy sensitivity $\epsilon_0$. In fact,  according to Eq. (\ref{eq:shhT}), the readout contributed noise grows linearly with $\epsilon_0$ and we find that a value  $T\phi  < 10^{-8}\epsilon_0\;$ K  is needed to limit the thermal noise contribution within the 20\% of the readout noise. On the contrary the optimal value of the noise stiffness $\kappa_0$ is mainly determined by the transfer function $T_{BA}(\omega)$ and remains unchanged.

\begin{figure}[ht!]
\includegraphics[width=8.6cm,height=5.cm]{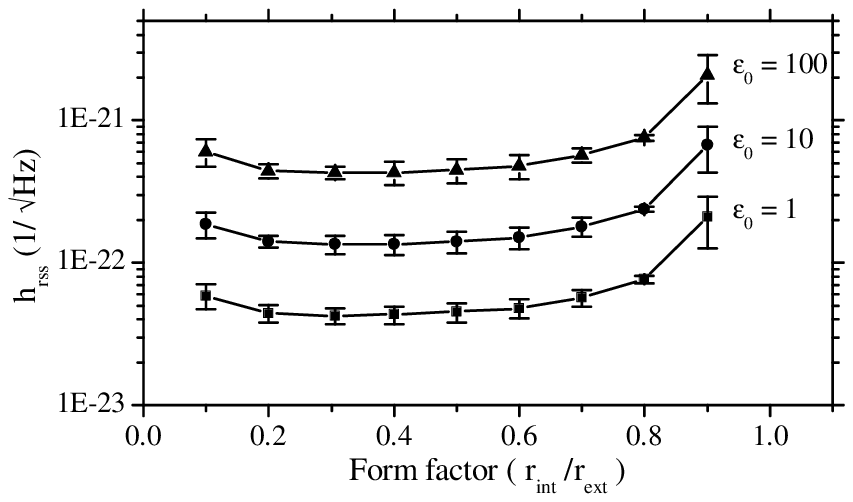}
\caption{\label{fig:ropti}  
The detector performances  for different values of the readout  energy sensitivity  is averaged over the 4 test frequencies and shown as a function of the form factor $r_{in}/r_{ext}$. The error bars indicate the   scattering of the sensitivity around the average value, a measure of the uniformity of the performances over the full bandwidth. A dissipation $T\phi  = {10^{-8}}{\epsilon_0}$ K  was assumed.
}
\end{figure}

\begin{figure}[ht!]
\includegraphics[width=8.6cm,height=5.cm]{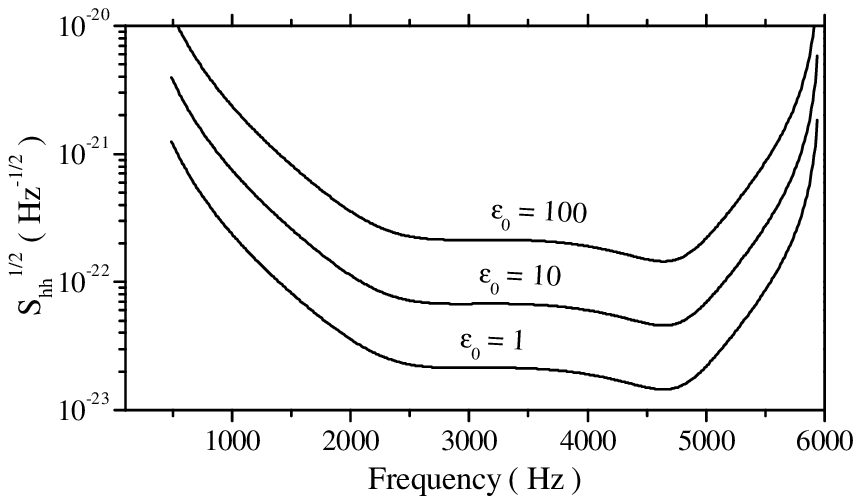}
\caption{\label{fig:shh} Predicted spectral strain sensitivities of the molybdenum detector: weight 22 ton, outer diameter 1 m, inner diameter 0.3 m and height 3 m. The curves  at different values of the readout energy sensitivity $\epsilon_0$ are evaluated with optimal values of noise stiffness $\kappa_0=1.7 \times 10^{11}\;$N/m and with dissipation  $T\phi  = {10^{-8}}{\epsilon_0}$ K. The corresponding displacement noise power of the optimal readout is $S_{xx}={6\times 10^{-46}}\,\epsilon_0\;$m$^2$/Hz and back action force power noise of $S_{ff}\simeq 1.8 \times 10^{-23}\,\epsilon_0\;$N$^2$/Hz. The curve with $\epsilon_0=1$ represents the detector SQL.}
\end{figure}

Figure \ref{fig:ropti} shows the detector performances at the optimal noise stiffness  averaged over the 4 test frequencies and then plotted against the form factor. Different values of energy sensitivity were considered with their corresponding optimal value of  $T\phi$. The noise stiffness was optimized for each point on the graph  and found essentially constant $\kappa_0\simeq 1.7 \times 10^{11}$ N/m.  The dependence on the inner radius is not strong, but  the best sensitivity is obtained for $0.2<r_{in}/r_{ext}<0.5$, that is when the antiresonance is located approximatively in the middle of the bandwidth. 
The GW sensitivity $S_{hh}$  of this optimal detector (molybdenum, $r_{int}=0.15$ m, $r_{ext}=0.5$ m, height 3 m) is shown in Figure \ref{fig:shh}.

\section{Conclusions}
\label{sec:conclusions}
In this paper we discuss wide-band (ie not-resonant) acoustic GW detectors in the framework of the elastic theory of continuous bodies and we take fully into account the 3-dimensional properties of signal, test-mass and readout. Once the readout (ie its spatial weight function) is chosen, transfer functions are calculated for the GW tidal force, the back action noise pressure exerted by the readout and the thermal noise of the test-mass. The SQL sensitivity limit of acoustic GW detectors is then derived.

On the basis of this analysis we design a new kind of acoustic GW wide band detector: the single-mass Dual detector, formed by an hollow cylinder equipped with 2 sets of selective readout displaced $\pi/4$ apart. This detector would offer the same advantages of a Dual detector, namely the kHz-wide frequency range of sensitivity and the back-action reduction effect: moreover we show that such a detector has no blind directions for impinging GWs, being in this sense omnidirectional even if not isotropic. These findings are analytically derived in the plain strain approximation and then extended to the full 3-dimensional body by means of FEM numerical analysis. We also show that the single-mass Dual detector can measure simultaneously both the amplitude and the polarization of the impinging GW. From the practical view-point the single-mass Dual detector removes the technological challenge of nesting two independent resonators which is one of the key issues of the recently proposed Dual detectors.

The general treatment of a GW acoustic detector is then applied to the single-mass Dual detector in an optimization process that gives the mass dimensions and readout noise properties resulting in the best sensitivity in the 2-5 kHz frequency range. We show that a molybdenum hollow cylinder single-mass Dual (outer radius $r_{ext}$ = 0.5 m, height $L$ = 3 m, form ratio $r_{int}/r_{ext}$ = 0.3 and readout angular aperture $\alpha=0.2\;$rad) can show a sensitivity at the level of $1\div 2 \cdot 10^{-23}/\sqrt{Hz}$ in the frequency interval 2 $\div$ 5 kHz and truly free from noise resonances, provided that $T\phi < 10^{-8}\;$K, as already achieved for this material.

The non-resonant readouts for a Dual detector would be evolutions both conceptually and in technology of the resonant readouts used in the GW bar detectors: the optomechanical one \cite{ottico}, based on Fabry-Perot cavities, and the capacitive one \cite{auriga2}, based on SQUID amplifiers. The main improvements needed to make these readouts suitable in a Dual detector configuration can be summarized as:
\begin{description}
\item[The wideband sensitivity.]
The  deformation of the test mass which forms the detector needs to be measured by devices that do not show mechanical resonances in the kHz wide frequency interval of high sensitivity.  
\item[The selectivity.]
The Dual detector properties can be  fully exploited by a readout scheme which is geometrically selective to the fundamental quadrupolar modes. This selectivity is also effective in reducing down to a negligible level the acoustic modes not sensitive to GW that would appear in the detector bandwidth. Moreover the readout system must sense the deformation of the resonant masses on a wide surface, in order to be less sensitive to the acoustic modes resonating at higher frequencies, which do not carry any gravitational signal. In this way the thermal noise of the detector is minimized, while preserving the sensitivity to the signal.  
\end{description}

Without the resonant amplification stage, the displacement sensitivity of readout employed in running GW bar detectors ranges in the $ 10^{-19}\;$m/$\sqrt{Hz}$ decade over a bandwidth of  hundred of Hz centered at about 1kHz. In order to meet the requirement of a Dual detector, an improvement is needed by about a factor of $10^4$ and these performance must be extended to cover the wider Dual bandwidth. For passive readouts, as the capacitor-Squid system used in AURIGA, a factor of $10^3$ improvement in the displacement sensitivity is foreseeable after specific R$\&$D devoted to increase the bias field (presently limited to about 1\% of the material breakdown electric field \cite{diamond}) and to increase the capacitor area (the transducer efficiency scales as the square root of the capacitance) and after a reduction of the operating temperature down to the 0.1$\;$K range (the SQUID amplifier additive noise is linearly proportional to the thermodynamic temperature). For active readouts as the optomechanical one \cite{ottico} the $10^{-23}\;$m/$\sqrt{Hz}$ range is reachable in principle by increasing both the finesse (up to $10^6$) and the light power (up to a few Watts); however such figures could end up not to be compatible with a cryogenic operation. Therefore for both technologies we can consider as reasonable goal a sensitivity at the level of $10^{-22}\;$m/$\sqrt{Hz}$ over the bandwidth of a Dual detector. A factor of 10 further could be gained via broadband mechanical lever amplifiers, as currently under investigation.

The wide-area and selectivity requirements could be easily satisfied by capacitive readouts. In fact they are currently implemented on surfaces of 0.05 m$^2$, and many capacitors could be properly connected to obtain the required total surface and selectivity, as shown in ref. \cite{bonaldi1}. For the optomechanical readout the implementation of selectivity  based on crossed cavities is under study; in order to meet the requirement of the wide area some of us has proposed a new configuration of Fabry-Perot optical resonator, the Folded Fabry-Perot \cite{ffp}, and its experimental investigation has already started \cite{marin}.

\section{Acknowledgements}
We thank Francesco Marin and Antonello Ortolan for fruitful discussions.

This work was supported by European Community (project ILIAS, c.n. RII3-CT-2004-506222), Provincia Autonoma di Trento (project QL-READOUT), European Gravitational Observatory (R\&D grant ``Wide band transducer for Dual detector'').

\appendix
\section{Modal analysis}
\label{appendix:modal}
\begin{figure}[t!]
\includegraphics[width=8.6cm,height=8.6cm]{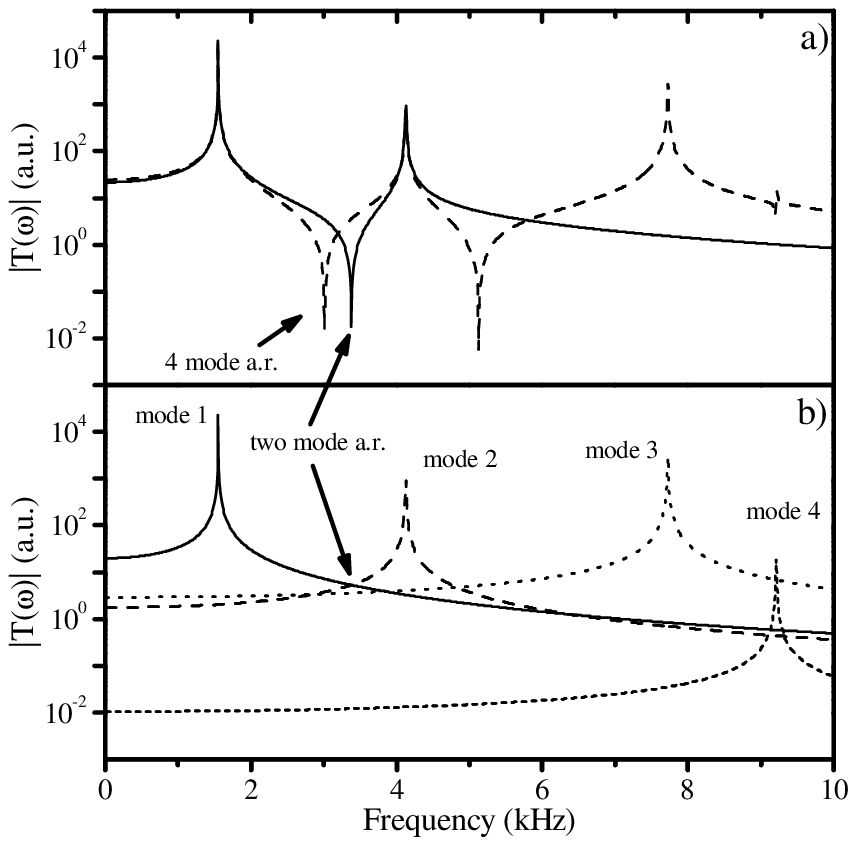}
\caption{\label{fig:tba}  
Typical transfer function for a 4 modes system where the driving force is applied with the same spatial weight of the readout. The resonance-antiresonance behavior is due to the superposition of the single mode responses and, in the hypotheses of low modal density and small damping, an antiresonance is invariably present between the two resonances.  (a) Continuous line:  transfer function obtained when the \textit{mode 1} and \textit{mode 2} contributions are only considered. Dashed line:  transfer function obtained from the contribution of all the 4 modes; the position of the first antiresonance is changed by the presence of the higher frequency modes contributions. (b) Transfer functions of each mode.
}
\end{figure}

The complex behavior observed in  a typical transfer function (Fig. \ref{fig:tba}a) 
is not surprising, as it is well known that structures with  low modal density \cite{modaldensity} exhibit a pronounced resonance-antiresonance behavior in their transfer function \cite{ewins}. 
Antiresonances provide important information on the dynamic behavior of the structure under consideration and have been applied to several engineering problems, as  reported in \cite{nam} and Refs. therein.

Considerable insight may be gained by considering the origin of  antiresonances. Within the modal expansion model, the system transfer function can be written in the form:
\begin{equation}
 {T}(\omega)=\,\sum_{n}\frac{
\,C_n}{(\omega^2_{n}-\omega^2)+ i D_{n}(\omega)}\,\:,\label{eq:modalconstant}
\end{equation}
where the explicit form of the modal constants $C_n$ and the dissipative factor $D_n$ can be 
obtained by a straightforward comparison with Eq. (\ref{genericresponse}).
As usual the resonant frequencies represent the zeros of the denominator polynomial, that in the case of small damping are approximatively given by the eigenvalues $\omega_n$. The eigenvalues  are  determined by the mechanical properties of the system and by the boundary conditions applied when solving Eq. (\ref{eigenproblem0}). The antiresonance frequencies are  the zeros of the numerator polynomial resulting from the sum, and their position on the frequency axis depends on the values of the modal constants $C_n$. 
In Figure \ref{fig:tba} we show $|\,{T}(\omega)\,|$ as sum of the contributions of each resonant mode  for a 4 mode system with $C_i>0$.

Some of the important features of the transfer function can be understood by using a simpler example with just two modes and no dissipation: 
\begin{equation}
{T}(\omega)=\frac{C_1(\omega^2_{2}-\omega^2)+C_2(\omega^2_{1}-\omega^2)}{(\omega^2_{2}-\omega^2)(\omega^2_{1}-\omega^2)}\,.
\end{equation}
If both modal constants are positive, a zero of the numerator polynomial $\omega_{ar}$ is invariably present in the range $[\omega_1,\omega_2]$ between the two resonances, at a frequency depending on the relative weight between the modal constants $C_1$ and $C_2$:
\begin{equation}
\label{eq:antiresonance}
\omega_{ar}=\sqrt{\frac{\omega_1^2+\frac{C_1}{C_2}\,\omega_2^2}{1+\frac{C_1}{C_2}}}\,.
\end{equation}

Positive modal constants appear in all  transfer functions where the driving force is applied with the same spatial weight of the readout, as for the back action transfer function $T_{BA}$ [Eq. (\ref{eq:TFbackaction})]. In this case there must be an antiresonance following a resonance, without exception and regardless of the complexity of structures \cite{ewins}. 
On the other hand, modal constants with opposite signs result in a smooth minimum between the two resonances, located at:
\begin{equation}
\omega_{min}=\sqrt{\frac{\omega_1^2}{1+\sqrt{-\frac{C_1}{C_2}}}+\frac{\omega_2^2}{1+\sqrt{-\frac{C_2}{C_1}}}}\,.
\end{equation}
Modal constants of the detector transfer function to GW [Eq. (\ref{gwtransfer})] can be negative or positive, depending on the integrals performed over the modes. In this case the sign is essentially determined by the phase relation between a specific mode and the GW excitation force, and by the phase relation between this mode and the weight function. Then we expect a mixture of antiresonances and smooth minima. 

We point out that this simple model is only indicative: the actual position of  antiresonances and smooth minima between resonances depends on the contribution of the other modes, as shown in Figure \ref{fig:tba}.


\begin{thebibliography}{<100>}
\bibitem{louisiana1}ALLEGRO http://gravity.phys.lsu.edu
\bibitem{auriga1}AURIGA http://www.auriga.lnl.infn.it
\bibitem{rog1}EXPLORER-NAUTILUS 

http://www.lnf.infn.it/esperimenti/rog/
\bibitem{rog2}P. Astone \textit{et al.}, Phys Rev. Lett. \textbf{91},  111101
(2003).
\bibitem{auriga2} L.Baggio \textit{et al}, Phys Rev. Lett.  \textbf{94}, 241101 (2005).
\bibitem{Thorne_300y}
K.~S. Thorne,  in {\em Three Hundred Years of  Gravitation}, edited
by S.~W. Hawking and W. Israel (Cambridge University Press,
Cambridge, 1987).
\bibitem{cerdonio1} M. Cerdonio, L. Conti, J.A. Lobo, A. Ortolan, L. Taffarello, and J. P. Zendri,  Phys. Rev. Lett. \textbf{87} 031101 (2001).
\bibitem{bonaldi1}M. Bonaldi, M. Cerdonio, L. Conti, M. Pinard,
G. A. Prodi, L. Taffarello, and J. P. Zendri, Phys. Rev. D \textbf{68}
102004 (2003).
\bibitem{shibata} M. Shibata, K. Taniguchi and K. Uryu,  Phys. Rev. D \textbf{71}
084021 (2005).
\bibitem{pretorius} F. Pretorius, Phys. Rev. Lett. \textbf{95} 121101 (2005).
\bibitem{campanelli} M. Campanelli \textit{et. al.}, gr-qc 0511048.
\bibitem{saulsonlibro}P. Saulson, \textit{Fundamentals of interferometric gravitational wave detectors}, (World Scientific, 1994).
\bibitem{Love} A.E.H. Love, \textit{A treatise on the mathematical theory
of elasticity} (Dover, New York, 1944).
\bibitem{Nmodes} The number of modes which must be considered  in order to obtain the displacement fields within a given approximation must be carefully evaluated and verified by   convergence tests.
\bibitem{meirovitch} L. Meirovitch, \textit{Methods of analytical dynamics}, (McGraw-Hill, New York, 1970).
\bibitem{saulson}P.R. Saulson, Phys. Rev. D \textbf{42}, 2437 (1990).
\bibitem{yamamoto1} K. Yamamoto \textit{et al.}, Physics Letters A \textbf{321}, 79 (2004).
\bibitem{ottico} L. Conti \textit{et al.}, J. Appl. Phys. \textbf{93}, 3589 (2003).
\bibitem{capacitivo} A. Marin \textit{et al}., Class. Quant. Grav. \textbf{19} 1991 (2002).
\bibitem{misner} C.W. Misner, K.S. Thorne and J.A. Wheeler, \textit{Gravitation}, (W.H. Freeman and Company, S. Francisco, 1973). 
\bibitem{braginsky} V.B. Braginsky  and  F.Ya. Khalili, \textit{Quantum Measurement}, (Cambridge University Press, Cambridge, 1992).
\bibitem{callen51}H. B. Callen and T. A. Welton, Phys. Rev. D \textbf{83},  34 (1951).
\bibitem{levin} Yu. Levin, Phys. Rev. D \textbf{57}, 659 (1998).
\bibitem{gazis}  D.C. Gazis, J. Acoust. Soc. Am. \textbf{30}, 786-794 (1958).
\bibitem{briant}T. Briant \textit{et al.}, Phys. Rev. D \textbf{67} 102005 (2003).
\bibitem{ansys} ANSYS, Inc., Southpointe, 275 Technology Drive,
Canonsburg, PA 15317; http://www.ansys.com
\bibitem{yamamoto}K. Yamamoto \textit{et al.}, Physics Letters A \textbf{305}, 18 (2002).
\bibitem{forward} R.L. Forward, Phys. Rev. D \textbf{17}, 379 (1978).
\bibitem{duffyMo} W. Duffy, Jr., J. Appl. Phys. \textbf{72}, 5628 (1992).
\bibitem{diamond}W. T. Diamond, J. Vac. Sci. Technol. A \textbf{16}, 707 (1998).
\bibitem{ffp}F. Marin, L. Conti and M. De Rosa, Phys. Lett. A \textbf{309}, 15 (2003).
\bibitem{marin}F. Marin, private communication.
\bibitem{modaldensity} In a dynamical system with   low modal density, the resonant frequencies of the modes are separated by a frequency range much greater than their bandwidth. 
\bibitem{ewins}D. Ewins, \textit{Modal Testing: Theory and Practice}, (Research Studies Press, Baldock, 2000).
\bibitem{nam} D. Nam \textit{et al.}, International Journal of Solids and Structures \textbf{42}, 4971 (2005).

\end{thebibliography}
\end{document}